\documentclass[sigconf,9pt]{acmart}
\copyrightyear{2026}
\acmYear{2026}
\acmBooktitle{DAC '26: Design Automation Conference} 

\acmConference[DAC '26]{}

\usepackage{hhline}
\usepackage{colortbl}
\usepackage{centernot}
\usepackage{pbox}
\usepackage{amsmath}
\usepackage{amsfonts}
\usepackage{url}
\usepackage{bm}
\usepackage{comment}
\usepackage{graphicx}

\usepackage{epstopdf}
\usepackage{multirow}
\usepackage{color}
\usepackage{tabu}
\usepackage{mdframed}
\usepackage{syntax}
\usepackage{fancyvrb}
\usepackage{subfig}
\usepackage[table]{xcolor}

\usepackage[linesnumbered,ruled,vlined]{algorithm2e}

\usepackage{soul}

\usepackage{paralist}
\usepackage{stmaryrd}
\usepackage{tikz}
\usepackage[referable]{threeparttablex}
\usepackage{tabularx}
\usepackage{booktabs}
\usepackage{array}
\usepackage{fancyhdr}
\usepackage{float}
\usepackage{xspace}
\usepackage{pifont}
\usepackage{balance}
\usepackage{xcolor}

\usepackage{tablefootnote}
\usepackage[normalem]{ulem}

\usepackage{tikz}
\usepackage{xcolor}

\newfloat{figtab}{htb}{fgtb}
\makeatletter
  \newcommand\figcaption{\def\@captype{figure}\caption}
  \newcommand\tabcaption{\def\@captype{table}\caption}
\makeatother

\usepackage{cleveref}
\crefname{figure}{fig.}{figures}
\Crefname{figure}{Fig.}{Figures}
\Crefname{table}{TABLE}{Tables}

\definecolor{princetonorange}{RGB}{255,143,0}

\definecolor{lightgreen}{RGB}{198, 224, 183}

\definecolor{lightred}{RGB}{240, 205, 176}

\newcommand{\TITLE}{FSGen\xspace}

\copyrightyear{2026}
\acmYear{2026}
\setcopyright{cc}
\setcctype{by}
\acmConference[DAC '26]{63rd ACM/IEEE Design Automation Conference}{July 26--29, 2026}{Long Beach, CA, USA}
\acmBooktitle{63rd ACM/IEEE Design Automation Conference (DAC '26), July 26--29, 2026, Long Beach, CA, USA}
\acmDOI{10.1145/3770743.3804052}
\acmISBN{979-8-4007-2254-7/2026/07}

\author{Jay Zhe-An Mok, Qijun Zhang, Zhiyao Xie\textsuperscript{*}}
\affiliation{%
  \institution{Hong Kong University of Science and Technology}
  \country{Hong Kong}
}
\affiliation{%
  \institution{\textnormal{\texttt{\{jzmok,qzhangcs\}@connect.ust.hk,eezhiyao@ust.hk}}}
  \country{}
}

\begin{document}



\title{DIM-SUM: Deep learning Inference Machine Specification and Unified Modelling, an open-source framework for generating customizable Sparse, Winograd, Systolic and Dense AI chips, with AI-driven fine-grained energy and power modelling.}
\title{Towards Netlist Foundation Model: A Self-Supervised and Cross-Stage-Aware Netlist Encoder via Text-Attributed Graph}
\title{Towards Netlist Foundation Model: A Multimodal Cross-Stage-Aligned Netlist Encoder via Text-Attributed Graph}
\title{Towards Netlist Foundation Model: \\ A Multimodal and Cross-Stage-Aligned Netlist Encoder \\ via Text-Attributed Graph}
\title{NetTAG: A Multimodal and Cross-Stage-Aligned \\ Netlist Foundation Model via Text-Attributed Graph}
\title{NetEncoder: A Multimodal RTL-and-Layout-Aligned \\ Netlist Foundation Model via Text-Attributed Graph}
\title{NetEncoder: A Multimodal and Cross-Stage-Aligned \\ Netlist Foundation Model via Text-Attributed Graph}
\title{DNN Inference Machine Specs to Unified Modeling}

\title{DIMSUM: DNN Inference Machine Specs to Unified Modeling, an Open-Source AI Accelerator Generator with AI-enhanced Fine-Grained Power Modeling}

\title{An Open-Source AI Chip Generator with a Comprehensive Space and an AI-enhanced Fine-Grained Power Modeling}

\title{DIMSUM: An Open-\underline{S}o\underline{u}rce AI Chip Generator with a Co\underline{m}prehensive \underline{D}esign Space and F\underline{i}ne-Grained Power \underline{M}odel}

\title{ Chi : An Open-\underline{S}o\underline{u}rce AI Chip Generator with a Co\underline{m}prehensive \underline{D}esign Space and F\underline{i}ne-Grained Power \underline{M}odel}

\title{ChiGong  : An Open-Source AI \underline{Chi}p \underline{G}enerator with a C\underline{o}mprehensive Desig\underline{n} Space and Fine-\underline{G}rained Power Model}

\title{An Open-Source AI Chip Generator with a Comprehensive Design Space and Fine-Grained Power Model\vspace{-.4in}}


\title{SpGen: Agile Fused and \underline{Sp}arse Spatial Array \underline{Gen}erator with Accurate Power Model for LLM Applications }

\title{FSGen: Agile \underline{F}used and \underline{S}parse Accelerator \underline{Gen}erator with Accurate Power Model for LLM Applications }




\author{
}


\pagestyle{empty}

\begin{abstract}

With the growing demand of artificial intelligence (AI) applications, large language models (LLMs) have become important workloads in many domains. The question of how to efficiently generate optimal AI chip accelerator designs remains unresolved and challenging. Currently, there is a lack of end-to-end design methodologies for efficient design space exploration (DSE). We propose \textit{\TITLE}, an agile framework for attention-based LLM accelerator generation with an early-stage PPA estimator. \TITLE supports fused operator dataflows and sparsity with a diverse design space and finds designs with 1.4$\times$ better power efficiency or 10$\times$ speedup with similar PPA metrics compared to prior work. Pareto-optimal designs have much better performance over a wide range of LLM benchmarks and have 58$\times$ better figures of merit (FoM). Design exploration is also faster due to our PPA estimators, which have better accuracy than prior art and reduce DSE runtime drastically. 
\end{abstract}


\maketitle


\renewcommand{\thefootnote}{\fnsymbol{footnote}}
\setcounter{footnote}{1}
\footnotetext{Corresponding author.} 



    
\section{Introduction}


Recently, artificial intelligence (AI) applications have grown rapidly, large language models (LLMs) have become important workloads in advanced chatbots, speech, visual, audio, and industrial automation domains. LLM inference on edge hardware has become an important application that is becoming more prevalent in embedded hardware in order to address issues of data security, network latency, and inefficiencies for cloud and server solutions~\cite{CambriconLLM}.

However, optimizing customized AI chip solutions for running LLM workloads remains a challenge. First, it remains challenging to define and characterize a comprehensive design space for hardware solutions. Second, due to the exponential design space and large design size of AI chips, design and optimization efficiency still remains a difficult problem. To alleviate the issue, end-to-end AI chip hardware generators for efficient design have been proposed by prior works. TensorLib/Rubick~\cite{Tensorlib,Rubick} generates systolic and multicasting spatial arrays for AI chip kernels. They propose a concise hardware design space and an exploration strategy that prunes invalid designs and enables efficient design exploration. BinaryLLM~\cite{BinaryLLM} uses low-bit quantization and generates systolic arrays for the LLM workload. They include KV cache support and a limited set of fused operations (i.e. matrix multiply + activation). They also propose an ML-guided design exploration strategy. Stellar~\cite{Stellar} is a generator framework for sparse systolic arrays for matrix multiplication kernels, supporting data compression and different fine-grain sparsity dataflows. 
However, current frameworks have several limitations as summarized below.




\textbf{Limited Design Space Configurability}. Important candidate designs are missing from prior hardware generators~\cite{BinaryLLM, Stellar}. 1) \textit{Fused operation dataflows}, which greatly reduce intermediate tensor storage by fusing multiple kernels and are often used for accelerating LLM operations, remain underexplored.
Dataflow is very important in the PPA trade-offs of customized AI chips~\cite{2019Understanding, TENET}.
2) \textit{Sparsity configurability} at both the algorithm and hardware implementation level is also limited. Sparsity is important for optimizing energy and performance. Prior work~\cite{Stellar} only supports one or a limited set of sparsity and sparse networks, and is unable to compare multiple types of sparsity in a unified framework.

\textbf{Insufficient Support to Implementation}. Many design exploration tools for LLM~\cite{LLMCompass, wu2022sparseloop} do not support automated RTL generation, preventing users from verifying their designs and quality of results (QoR) on FPGAs or as digital ASICs.

\textbf{Limited Early-Stage Estimator Accuracy Due to Simplification.}  Existing architecture exploration tools suffer from over-simplified power models. Several works assume that the per-operation power and energy are constant or limited to several states ~\cite{2019Understanding, 2019Accelergy,2014Aladdin}). In reality, the energy is highly sensitive to input activities.

To overcome these challenges, we propose \textit{\TITLE}, an open-source agile framework for attention-based LLM accelerator generation with early-stage ML-estimators\footnote{Code released at https://github.com/hkust-zhiyao/FSGen}. The main contribution of our work is a representation for sparse and fused operator dataflows that is more flexible and expressive of various LLM accelerators and enables efficient exploration of a larger design space compared with prior works. We support multi-level sparsity architectures, full configurability over fused operator tilings and loop orders, which allow more energy-efficient and high-performance designs to be found.
Our contributions are summarized below.

\begin{itemize}



     \item \textbf{A Comprehensive Design Space}.  \TITLE supports full configurability of fused operator dataflows and sparsity for LLM acceleration. Fused operator greatly reduces memory accesses and improves performance. We also propose the \textit{sparse set and tiling}, a unified representation which bridges the gap between high-level design and sparse hardware. 
     
    \item \textbf{Automated Generator from Design Space to Implementation}. A highly configurable end-to-end Chisel-based RTL generator, enabling agile implementation and verification.
    
    

    \item \textbf{Efficient Exploration of AI Accelerators}. \TITLE supports fast and accurate early-stage ML-based PPA estimators for power modelling. The model is fast to train, accurate, and generalizes well to full designs. Based on our generator and early-stage models, we enable efficient early-stage design space exploration and rapid implementation and verification of designs, reducing design time and difficulty. 
    
    
    
    
\end{itemize}

In the remaining sections, we introduce the background (Sec.~\ref{sec:background}), \TITLE's high-level representations (Sec.~\ref{sec:designspace}), hardware design space (Sec.~\ref{sec:hardwarespace}), hardware generator (Sec.~\ref{sec:generation}), and PPA estimators (Sec.~\ref{sec:PPA}).

\begin{figure}[t!]
  \centering
   \vspace{-.45in}
  \includegraphics[width=1\linewidth]{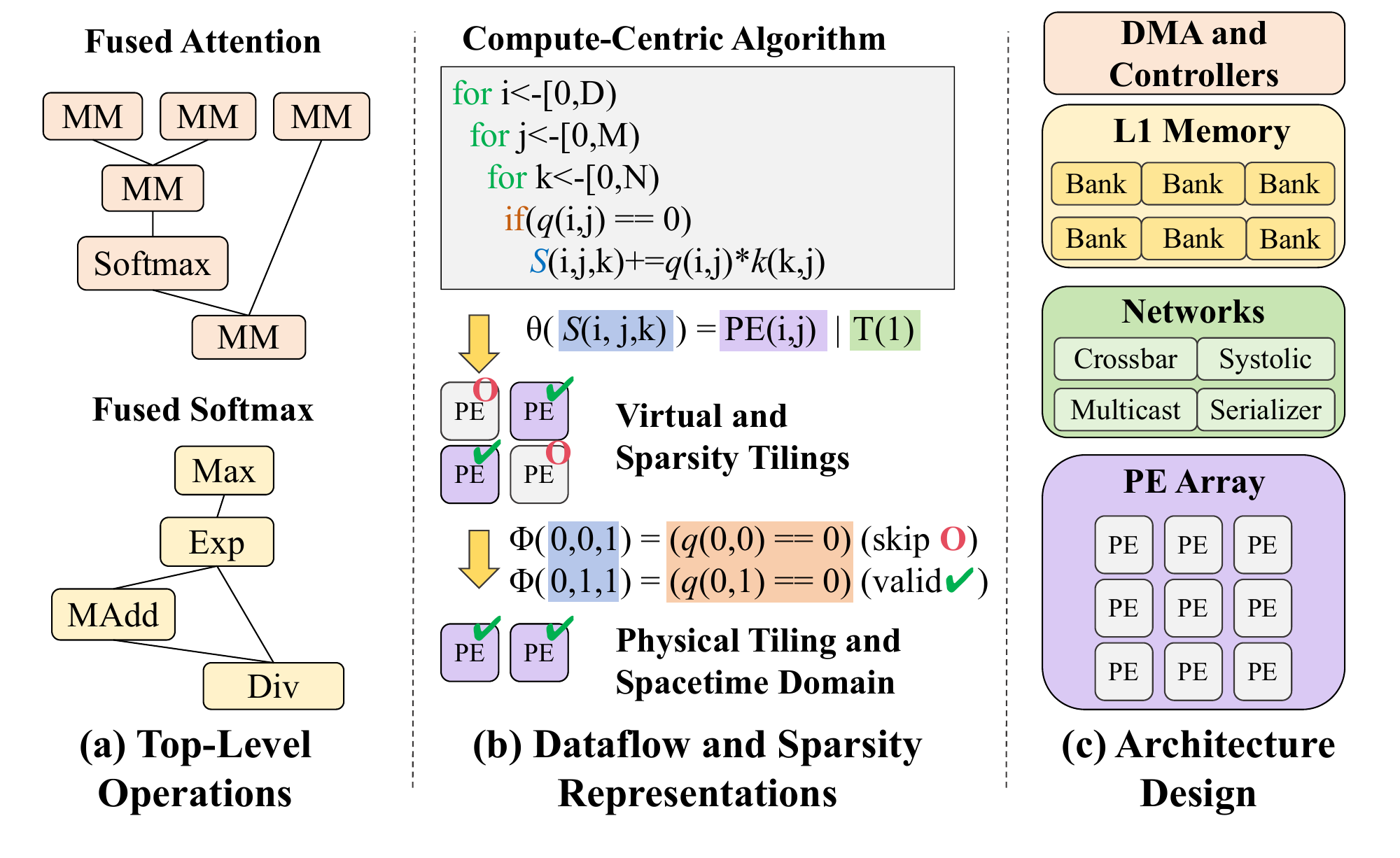}

 \vspace{-.1in}   
    \caption{Our proposed design space and generator covering (a) fused operations, (b) sparsity, and (c) architecture.   }
     
  \label{fig:teaser}
 \vspace{-.2in}
\end{figure}

\section{Background} \label{sec:background}

\textit{Hardware generators} are an agile design methodology for converting algorithms into hardware. Referring to~\cite{ TENET, Rubick}, algorithms are composed of loop instances and a tensor algebra. The \textit{iteration domain} $D_{S}$ is the set containing all loop instances $S$, which is a tensor algebra indexed by loop iterators $\vec{n}$. 
\vspace{-.05in}
\begin{align}
D_S =\{S(\vec{n}) | \vec{n} = (i,j,...)\}
\end{align}

A \textit{spatial array} consists of a processing element (PE) array, memory and interconnects. 
The \textit{space-stamp} is $PE(\vec{p})$ where $\vec{p}$ is a positional index of the PE array. The \textit{time-stamp} $T(\vec{t}),\vec{t}=(t_0, t_1, ..., t_n)$ corresponds to the schedule, where $t_0$ corresponds to the innermost loop and $t_n$ the outermost. Together, the space- and time-stamps form the space-time domain $D_{st}$.

A \textit{spatial dataflow} $\Theta$ is a mapping from a loop instance $S$ to a space- and time-stamp of a spatial architecture.
\begin{equation}
    \Theta_{D_S \xrightarrow{} D_{st}} =\{ {S(\vec{n}) \xrightarrow{} (PE(\vec{p}) | T(\vec{t})) \} }
\end{equation}
A \textit{fused operation dataflow}, known as kernel or operation fusion, is a topology on multiple dataflows. Fused operation allows intermediate tensors to be passed directly to the next PE array for processing, thereby reducing costly memory accesses~\cite{Chimera, fusedLayer}. Fig.~\ref{fig:teaser} (a) shows fused operator dataflows for attention and softmax. 


\section{Accelerator High-Level Representation }
\label{sec:designspace}
\TITLE's representation can characterize complex dataflows such as fused operators, i.e., multi-nested imperfect loops, which are more realistic of industrial designs and represent a larger variety of designs, ranging from softmax to fused matrix multiply chains, unlike prior works such as~\cite{Rubick, Tensorlib, Stellar} that only consider perfectly nested loops and single dataflows. In addition, unlike prior art that couples sparsity to dataflow~\cite{Stellar}, \TITLE's sparsity design spaces are defined mathematically, general and orthogonal to dataflow, aiding in efficient design exploration. We now discuss the high-level fused operator dataflow (Sec.~\ref {sec:otherfused}) and sparsity representations (Sec.~\ref {sec:sparity}).

\begin{figure}[t!]
  \centering
   \vspace{-.45in}
  \includegraphics[width=1\linewidth]{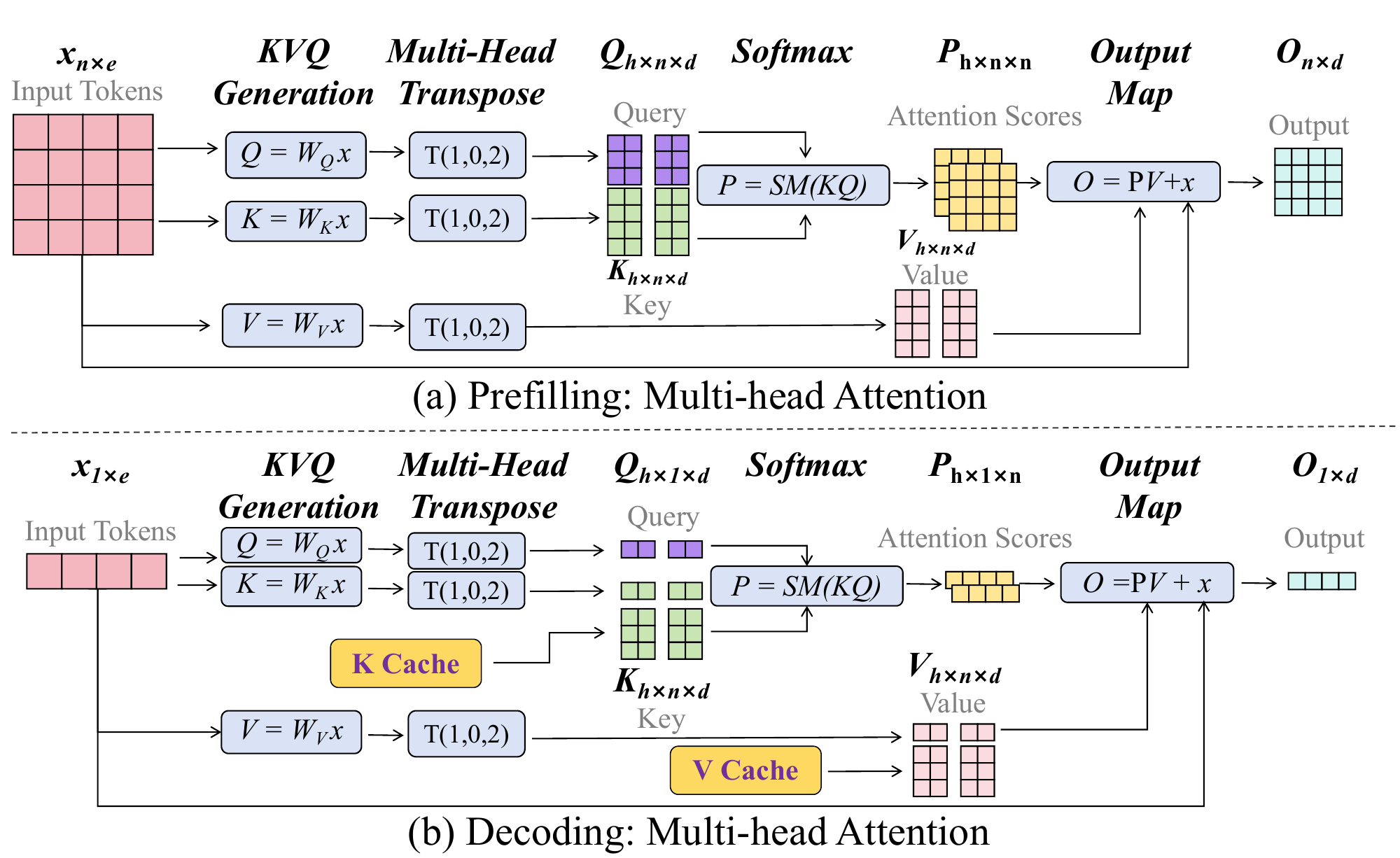}
   \vspace{-.25in}
  \caption{Typical attention flows. (a) Prefill-Stage Multi-Head Attention (b) Decode-Stage Multi-Head Attention.
}
  \label{fig:algorithm}
 \vspace{-.2in}
\end{figure}


\subsection{Fused Operator Dataflows for LLM}
\label{sec:otherfused}

Fused operator dataflows are equivalent to multiple perfect loop nests sharing a common set of loop variables and iterators. We apply fused operator dataflows to important LLM layers, such as attention, feed-forward networks (FFN), KVQ generation, and normalization layers. Referring to Fig.~\ref{fig:algorithm}, multi-head attention (MHA) consists of several tensor algebras. The fused operator iteration domain for the prefill stage is represented as several loop instances $S_i$, with iterators $\vec{n} = (b, n, m, h,d,e,o)$, which are the batch, query, key/value sequence, head, hidden size, and input/output embedding, respectively, and SM refers to softmax. The fused operator chains the softmax and output operations to the attention score, targeting the main bottleneck. $S_1,S_2,S_3$ represent score calculation, softmax and the output matrix multiplication, respectively. The iterators and shapes can be inferred from Fig.~\ref{fig:algorithm}. 
\begin{flalign}
 \begin{split}
 S_1: S \mathrel{+}= Q \times K ;\ \ S_2: P = \text{SM}(S); \ \ S_3: O = V \times P + x \\
  \end{split}
\end{flalign}

The decoding stage is similar. The key difference is that only 1 token is generated, and prior key-value tokens are loaded from the KV cache.
The complete fused operator dataflow is a series of dataflows for each iteration domain $D_{S_i}$ and PE array $PE_{i}$,
\begin{flalign}
\begin{split}
    &\Theta_i =\{ {S_i(\vec{n}) \xrightarrow{} (PE_i(\vec{p}) | T_{i}(\vec{t})) \} }\\
\end{split}
\end{flalign}


Group-query attention (GQA) is very similar to MHA, but with the query partitioned into groups $g$ that share the same KV heads across all groups. 
Other layers are similarly represented.

\subsection{Sparse Dataflows}
\label{sec:sparity}

We now discuss our sparsity representation. Our framework can easily represent sparse combinations and group sparsity, which are not easily expressed by sparse data structures from prior works~\cite{wu2022sparseloop, Stellar}. Our sparse dataflows are also more general and flexibly configured, improving the design space.



\textbf{Sparse Set and Mapping}. The \textit{sparse set} $Q$ is a set of \textit{sparse mappings} $\Phi(F(\vec{n})) $, which map from iterators $\vec{n}$ and tensors $\vec{x},\vec{y} ...$ to the Boolean domain $\mathbb{B}^{F(\vec{n})} = \{0, 1\}^{F\vec(n)}$, where $F(\vec{n})$ is the cartesian product of a subset of iterators. 
\begin{equation}
Q_{ D_S \times (D_x, ...) \xrightarrow{} \mathbb{B}^{F(\vec{n})}} = \{ \Phi[\vec{n}, \vec{x} , ... ](F(\vec{n})) \ |\  \vec{n},\vec{x} \in D_x, ... \}
\end{equation}


We highlight several examples of sparse sets $Q$ used in our work. Weight sparse key generation as in~\cite{FTRANS} is represented as $Q_w = \{\Phi_w(\vec{n}) = (W_K(e,d) == 0)\}$, where $W_K$ is the key generation weights. Window attention as in~\cite{LongFormer} is $Q_{win}=\{ \Phi_{win}(\vec{n}) = (n - m \leq W) \& ( n - m \geq 0) \}$, where $W$ is the window length, $n$ and $m$ are the attention query and key's sequence length dimensions, respectively. For query-key pruning as in~\cite{Spargeattn}, given $mm,nn$ secondary tilings, P is average pooling, omitting the batch, head and hidden size iterators for simplicity,
 \begin{equation} 
 Q_g=\{\Phi_g(nn,mm) = \text{TopK}( \text{P}(q[nn,n]) \times \text{P}(k[mm,m]) )\}
 \end{equation}
 Note that in the final example, the output domain is a subset of the full domain, namely ${F(\vec{n}) = nn \times mm} \subset \vec{n} = nn \times mm \times n \times m$. As a result, 
 we observe that a sparse mapping $\Phi$ is \textit{value-based}, if $F(\vec{n}) = \vec{n}$, which means fine-grain values determine skipping. Otherwise, it is \textit{group-based} because a group of tensors are skipped. 





\begin{figure}[t!]
  \vspace{-.45in}
  \centering
  \includegraphics[width=1\linewidth]{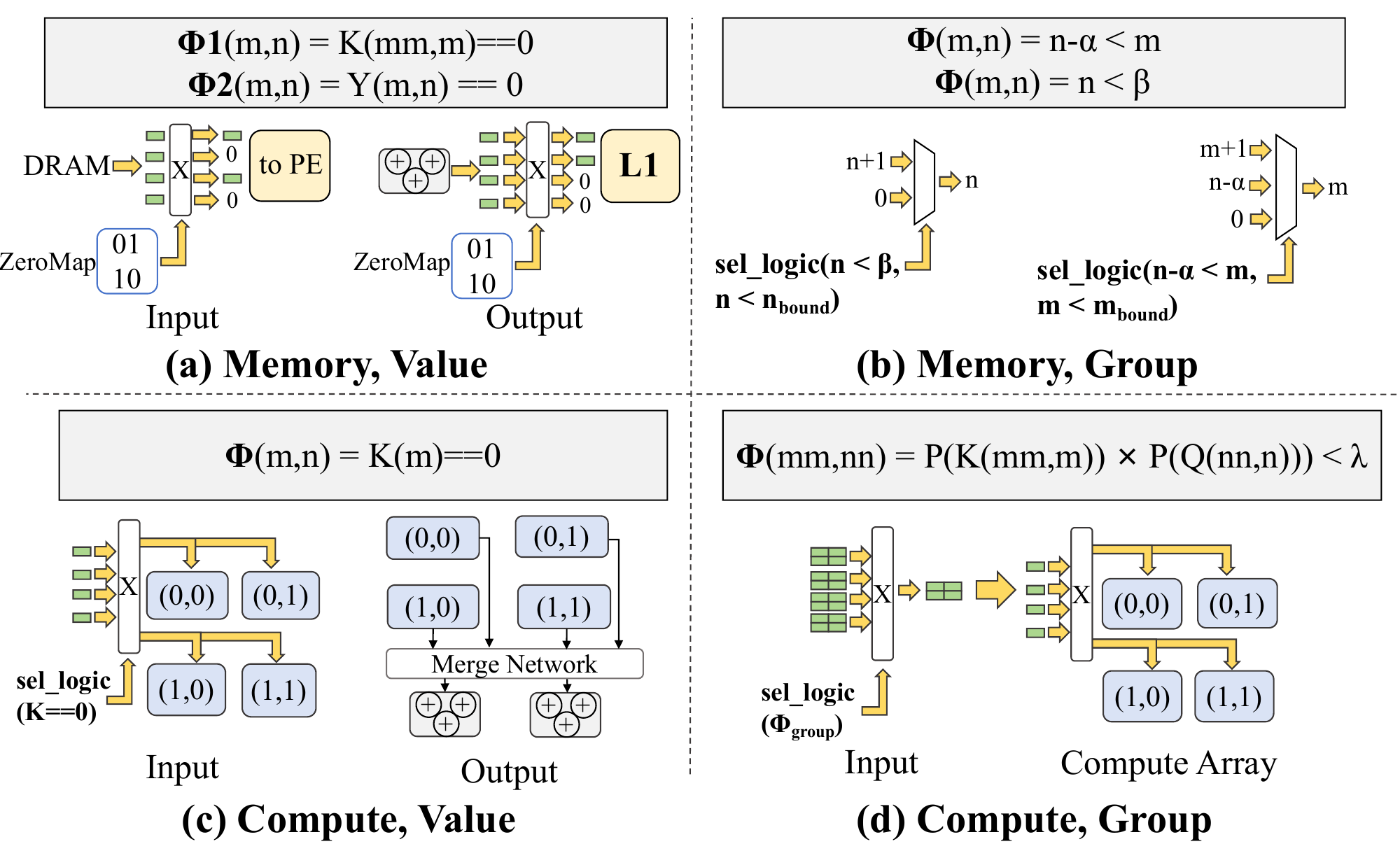}
     \vspace{-.3in}
  \caption{Different types of sparsity from sparse mappings. }

  \label{fig:sparseset}
 \vspace{-.2in}
\end{figure}

\textbf{Sparse Dataflow.} \textit{Compute- and memory-based sparse dataflows} are bindings between $Q$ and a dataflow $\Theta$, or the DMA unit, respectively. Fig.~\ref{fig:sparseset} shows the main types of sparse dataflows, including the sparse mapping $\Phi$ and its location. Sparse dataflow bridges the gap between the high-level design space and the underlying hardware.

\section{Hardware Design Space} 
\label{sec:hardwarespace}

The hardware design space is the set of mappable designs from the high-level representations. We present several methods that benefit performance and support design spaces that are not well supported by previous generators~\cite{BinaryLLM,Stellar}. We now discuss fused operator (Sec.~\ref {sec:hardwarespatial}) and then sparsity hardware spaces (Sec.~\ref{sec:hardwaresparse}).


\subsection{Fused Operator Design Space}
\label{sec:hardwarespatial}


Fused operator dataflows are composed of multiple dataflows, each is configured by a set of iterator tilings, loop order, and tensor accesses as determined by reuse analysis~\cite{Tensorlib}. Various data-stationary dataflows are set by the loop order ($\vec{t}_{loop} = (t_1, ..., t_n)$ time-stamps). We now discuss other important configurable spaces.





\subsubsection{  Hetereogenous Tiling Strategy } 
When tilings of different PE arrays are the same, as in~\cite{Spatten, A3}, downstream arrays are under-utilized because they need to wait for valid upstream data. To improve utilization, we propose a heterogeneous tiling strategy. From Fig.~\ref{fig:FusedOptim} (a), downstream tilings can be smaller, as long as the bandwidth is less than the upstream bandwidth. The maximum utilization is when the bandwidths are matched. 






\subsubsection{ Fused Operator Intermediate Tensor Cache}

\begin{figure}[t!]
  \vspace{-.45in}
  \centering
  \includegraphics[width=1\linewidth]{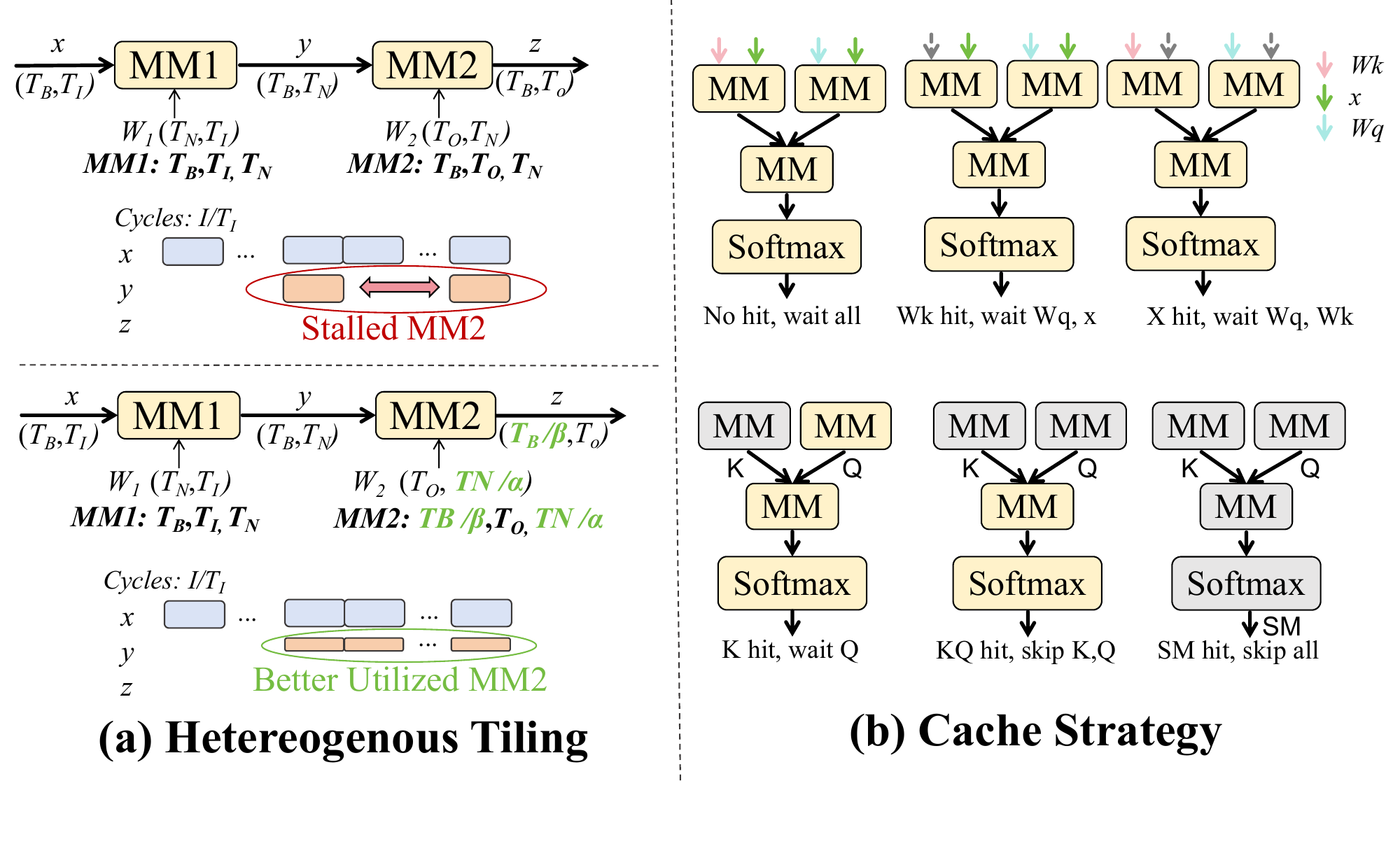}
     \vspace{-.4in}
  \caption{Fused operator dataflow optimizations. }

  \label{fig:FusedOptim}
 \vspace{-.2in}
\end{figure}

Cache buffer sizes for each tensor in the dataflow can be tuned. We propose a multi-tensor caching strategy. As shown in Fig.~\ref{fig:FusedOptim} (b) for an MHA dataflow, there are regular cache hit cases, and when both query and keys have cache hits, upstream arrays can be entirely skipped. 

\subsubsection{ LLM Dataflow Optimizations}
We implement two important optimizations targeting LLMs. First, to support both LLM stages, we modify the fused operator dataflow to be reconfigurable. The prefill mode is the same as the original dataflow. The decode mode allows key and values to be read from the cache, skipping extraneous KVQ generation. Second, we modify the softmax fused operator dataflow to perform online softmax~\cite{FlashAttention, Softermax}, effectively obviating a large softmax input cache and improving utilization. 

\begin{figure*}[t]
  \centering
\vspace{-.5in}
  \includegraphics[width=0.9\linewidth]{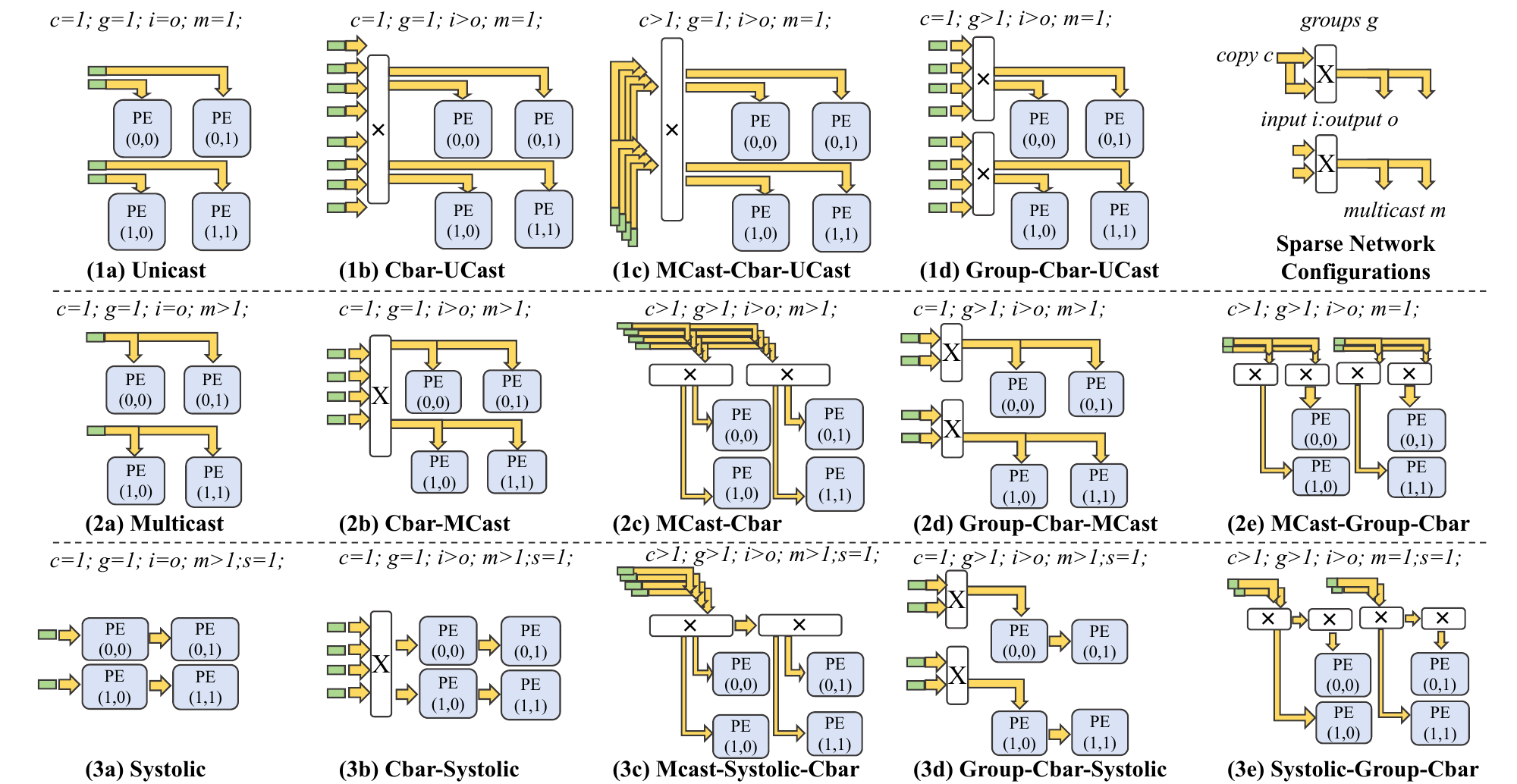}
       \vspace{-.1in}
       
    \caption{Design space of networks with a 2$\times$2 PE Array. The space is derived from sparse mappings $\Phi$ and sparsity tilings. }

  \label{fig:valsparse}
 \vspace{-.1in}
\end{figure*}


\begin{algorithm}[b!]

\caption{Compute-based Value Sparsity Mapping}
 
\label{algo:valmapping}
\SetKwInOut{Input}{Input}
\SetKwInOut{Output}{Output}

 \Input{
    High-Level Parameters ($\Phi$, $TT$, $T$, $S$, $\Theta$, $\vec{n}$), Tensor $x$.
    }
    \Output{
    Sparse Network Configuration for $x$ ($c,g,i,o,m$) 
    }

$A, E, I$ \hfill $\rhd$ Full, external, and inner iterators of a tensor
 
 $B_{x} = \{\vec{n} \}/ A_{x}$; $B_{\Phi} = \{\vec{n} \}/ A_{\Phi}$;  \hfill $\rhd$ Iterator complements
 
 

 
  \If {$A_{x} \in A_{\Phi}$} {
     c = 1;  g = $\prod\limits_{i}^{E_{\Phi} } {  \frac{ TT_i}{T_i} }$; m = $\prod\limits_{i}^{B_{\Phi} } TT_i$; i=$\prod\limits_{i}^{A_{\Phi} } \frac{S_i}{g}$; n = $\prod\limits_{i}^{A_{\Phi} } \frac{TT_i}{g}$;
    }
\vspace{-.05in}
 \Else{
     c = $S(B_x)$; g = $\prod\limits_{i}^{E_{x} } \frac{TT_i}{T_i}$; m = 1; i= $\frac{1}{g} \prod\limits_{i}^{E_{x} } TT_i \prod\limits_{j}^{I_{x} } S_j $; n =$\prod\limits_{i}^{A_{x} } \frac{TT_i}{g}$;
     
 }


\vspace{-.05in}
\end{algorithm}

\subsection{Sparsity Hardware Design Space}
\label{sec:hardwaresparse}

Sparse accelerators contain a PE array and several sparse networks, which determine the data selected into the PE array. In order to support both value and group-based sparsity, the sparse network is multi-level: the group-level network determines the coarse-grained group of tensors to enter the PE, and the value-level network then selects the fine-grained values. The sparse dataflow and sparse tilings define the hardware. A \textit{sparse tiling} encompasses the physical sub-tile $T$, physical tile $TT$, virtual tile $S$, and virtual group tile $SS$. Intuitively, $TT$ represents the PE array size, and $S$ and $SS$ the value- and group-level network, respectively. $T$ allows partitioning of the value-level network to improve timing. The compact notation for the sparse tiling of iterator $i$ is $i=T_i:TT_i:S_i:SS_i$. We now discuss the design space for each type of sparse mapping $\Phi$. 





\begin{figure}[b!]
\vspace{-.1in}
  \centering
  \includegraphics[width=1\linewidth]{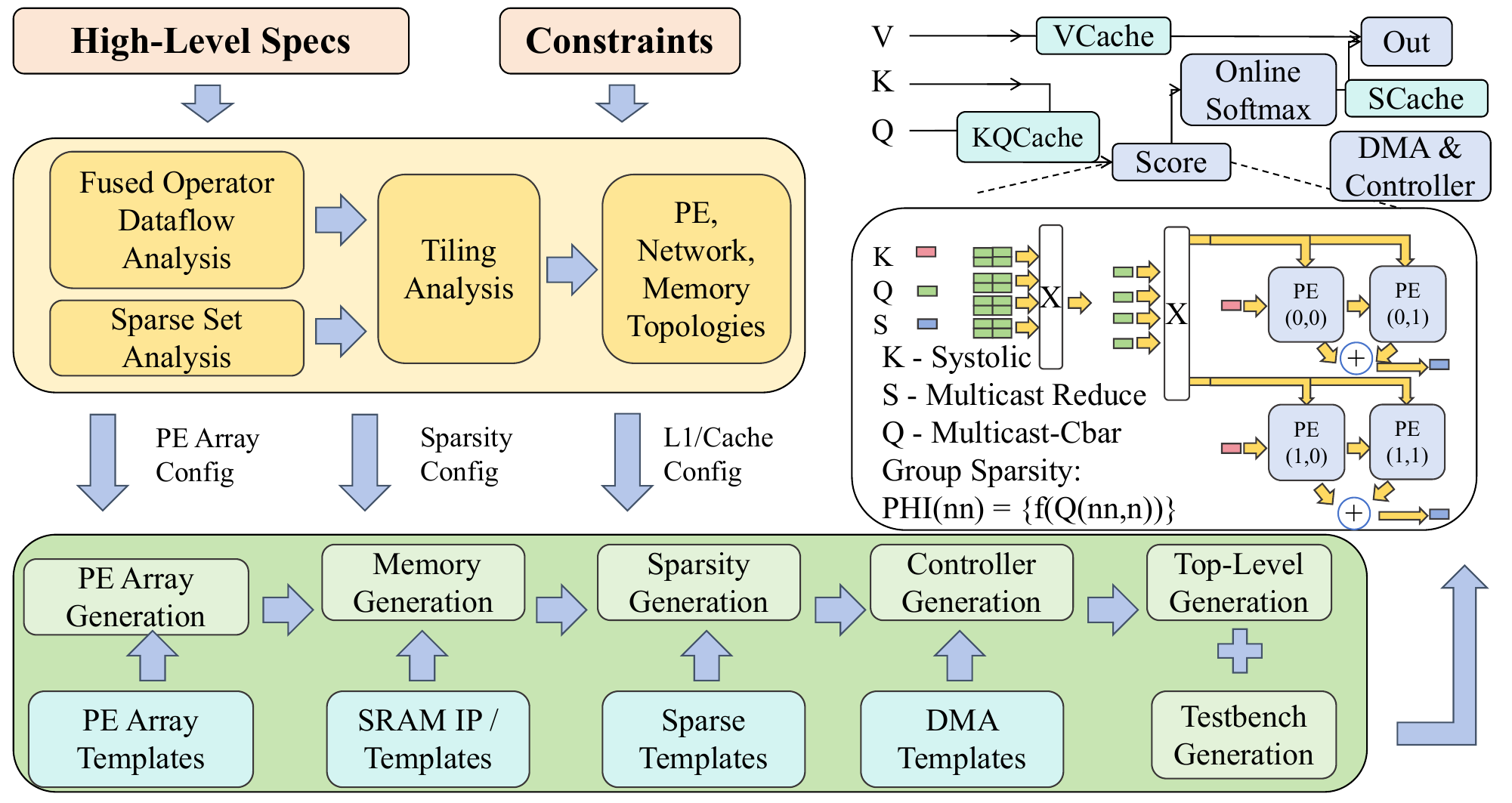}
   \vspace{-.25in}
  \caption{ Our hardware generation flow. High-level design parameters are mapped into RTL hardware. }
  \label{fig:generator}
\end{figure}

\subsubsection{Compute-Value Sparse Design Space}

Compute-value sparse mappings are mapped into sparse networks. \textit{Sparse networks}, or crossbars, are configured by the parameters copy \textit{c}, group \textit{g}, multicast \text{m}, input/output ports \textit{i} and \textit{o}, and systolic \textit{s}. As illustrated in the top right of Fig.~\ref{fig:valsparse}, \textit{c} is the fan-in of each input port, \textit{g} the number of crossbars, \textit{m} the fan-out of each output port, and \textit{s} the systolic movement as determined by reuse analysis similar to~\cite{Tensorlib}. Fig.~\ref{fig:valsparse} tabulates the space of sparse networks with a 2$\times$2 array. There are two cases for each parameter, i.e., $c,g,s,m=1$ or $>1$, $i=o$ or $i$ > $o$. Hence, there are $2^5 = 32$ network topologies for each tensor. Removing invalid and inefficient cases (copy and multicast both greater than one), as shown in Fig.~\ref{fig:valsparse}, there are a total of 14 networks: rows (1) are unicast, (2) multicast, and (3) systolic networks; columns (a) are non-sparse cases, (b) single networks, (c) grouped networks, (d) multicast networks, and (e) mixed combinations.

Algorithm~\ref{algo:valmapping} shows how high-level parameters are mapped to sparse network parameters. The key idea behind algorithm~\ref{algo:valmapping} is to use $TT$ and $T$ to configure the group, which reduces critical path delays, and $S$ and $TT$ to configure the ratio of the N:M fine-grained sparsity network to improve throughput. Internal iterators, such as the input channel of matrix multiply, do not need to be grouped because they are reduced. $s$ is determined by reuse analysis, as in~\cite{Tensorlib}. For example, for weight sparse matrix multiply with sparse tilings, $\Phi_w=(w==0)$, tensor $x$ maps to $c=\frac{TT_n}{T_n}, g=\frac{TT_b}{T_b},m=1,i=\frac{TT_b}{g S_i}$. Depending on the sparse tilings, the network can become grouped networks (c) or mixed combinations (e) because $m=1$. Output mergers are also added to collect outputs from different groups.







\subsubsection{Compute-Group Sparsity Design Space}

Compute-value group sparse mappings are configured by $S$ and $SS$, and use a similar algorithm to Algorithm~\ref{algo:valmapping} to map sparse networks. The main change is 1) the network's $i$ and $o$ correspond to $SS/S$ and $1$ instead of $S$ and $TT$, and grouping no longer exists because the sparse network is not affected by critical path delay issues, and 2) networks are placed before the value ones (or PE array), as shown in Fig.~\ref{fig:sparseset} (d).

\subsubsection{Memory-based Sparsity Design Space}

Memory-based sparse mappings change the state machine (FSM) in the DMA unit. Memory-value sparsity, otherwise called data compression, reduces memory calls by invalidating and skipping unnecessary off-chip requests. Memory-based group sparsity changes the main loop instance counter, which will increment to the specified bounds set by the sparse mapping's condition.

\begin{figure*}[t]

 \vspace{-.55in}
  \centering
  \includegraphics[width=1\linewidth]{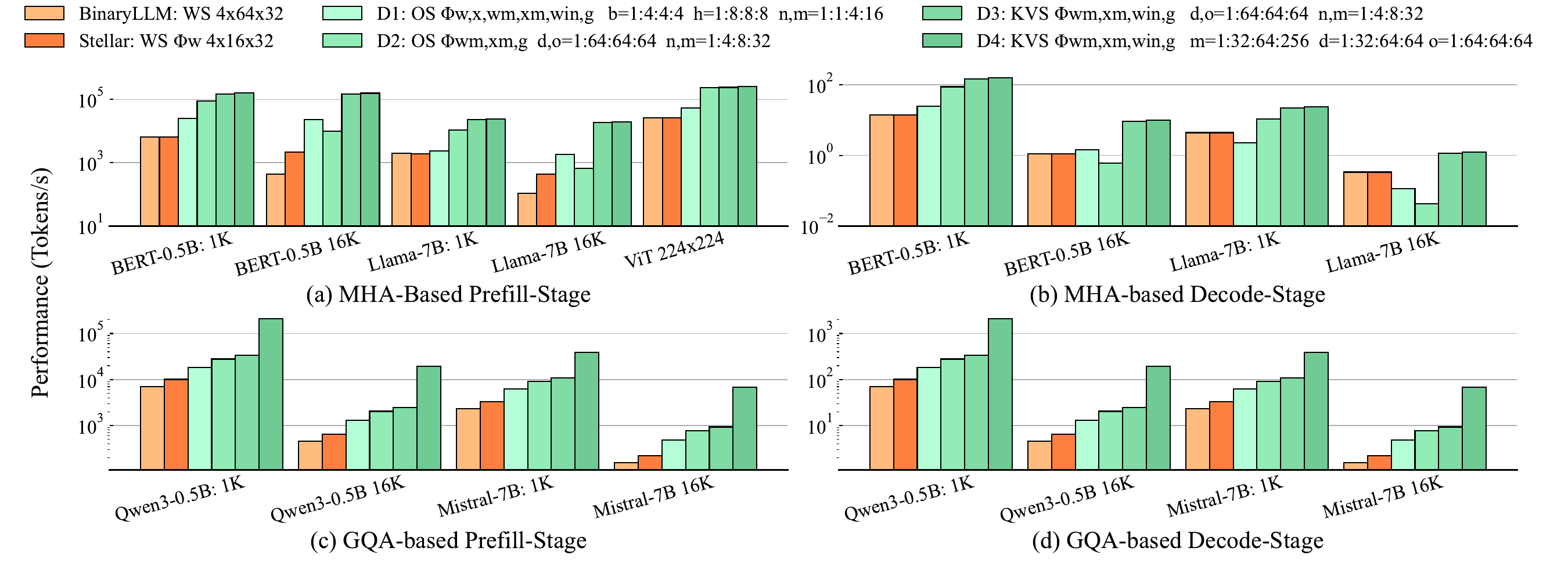}
   \vspace{-.25in}
  \caption{ Performance comparison of Pareto-optimal designs for a variety of LLM benchmarks and token lengths. }
  \label{fig:dense}
 \vspace{-.1in}
\end{figure*}

\section{Hardware Generation}
\label{sec:generation}
Our work generates attention-based LLM accelerator core RTL in an agile, automated and efficient manner using Chisel~\cite{Chisel}. Fig.~\ref{fig:generator} shows the flow. We highlight the steps used in hardware generation.

\textbf{High-Level Design Specification Input.} The input consists of the attention type, dataflow, and sparsity configurations. Fused dataflow parameters include loop order, tilings, and cache sizes. Sparsity parameters include the location, sparsity set and tilings.
 

\textbf{Design analysis and Hardware Output.} PEs are determined from the dataflow, similar to~\cite{Tensorlib}, but with queues added in between arrays and pipelines to support fused operation. Memory arrays are based on constraints and sized to store a virtual tile per cycle. Compute-based sparsity networks are based on Algorithm~\ref{algo:valmapping}. Memory-based sparsity modifies the DMA's FSM controller. The units are then connected to form the top-level module, which exposes an AXI programming interface. 

\textbf{Software-compiler and Testbench Support}. The compiler reads in LLM models, loads weights, and converts each layer into executable kernels. The testbenches support kernel-level simulation, e.g., attention, FFN; full-layer simulations execute multiple kernels. To support LLM operations not limited to attention, the architecture is \textit{reconfigurable.} For FFN layers, the softmax unit is swapped with an activation array; for matrix multiply, the score and output matrix multiply units are run in parallel, supporting a larger tiling. 



\begin{table}[b!]
\vspace{-0.15in}
\centering
\begin{tabular}{ c|  c|c|  c|c|  c|c }
\hline

 
Design &  tps ↑   & $mm^2$↓&  mW ↓  &  MHz ↑  & Util. ↑ &   FoM ↑ \\
\hline
\cellcolor[HTML]{ffbc7e} ~\cite{BinaryLLM}  &  1.9  & 19.3   & 2120 & \textbf{1000} & 58$\%$ &  0.05 ($0.3\times$)  \\
\cellcolor[HTML]{ffbc7e} ~\cite{Stellar}   &  1.8   & 10.4   &  1130 & 565 & 50$\%$  & 0.15 ($1\times$) \\
\cellcolor[HTML]{b4ffd8}  D1              &  2.3 & \textbf{3.58} & \textbf{374} & 812 &  \textbf{93$\%$} &  1.7 ($11\times$)  \\
\cellcolor[HTML]{b4ffd8} D2               &  10   & 3.60 & 377 &  712 & 71$\%$ &  7.3 ($48\times$) \\
\cellcolor[HTML]{b4ffd8} D3               &  22   &  4.79 & 509  & 823 & 39$\%$&  \textbf{9.0 ($58\times$)}  \\
\cellcolor[HTML]{b4ffd8} D4               &   \textbf{23}   & 7.08 & 763 &  811 &  39$\%$ & 4.2 ($28\times$)  \\
\hline
\end{tabular}
    \caption{PPA comparison of different designs. The FoM is tokens/s (tps) / power (W) / area ($mm^2$). The tps is an average of the MHA decode-stage benchmarks from Fig.~\ref{fig:dense}. } 
    \label{tbl:ppa}
    \vspace{-0.25in}

\end{table}
\section{Hierarchical Power Modelling}
\label{sec:PPA}

We now discuss our methodology for fast PPA modelling for our designs. Having an accurate early-stage model alongside a generator enables designers to quickly explore the exponential design space of designs rapidly and receive early-stage feedback.

Our early-stage power model is based on primitives. Primitives are the basic modules of accelerators, which include processing elements (i.e., adders, multipliers, exponent units), networks (i.e., crossbars, systolic pipelines, multicasts) and memory (i.e. SRAM banks, registers). The total power or area is therefore the sum of all primitives. The power model $f$ of each primitive is, with $P$ being the output power:  
 \vspace{-.1in}
\begin{flalign}
 \begin{split}
    P &= f(t, h, \alpha, \beta)
  \end{split}
\end{flalign}
Inputs include the primitive type $t$, parameters $h$, signal toggle $\alpha$ and zero-bit count $\beta$. $\alpha$ and $\beta$ are estimated from an early-stage cycle-accurate architectural simulator by sampling the executed dataflow and counting the signal bit-level toggles and zero-bit counts per cycle. We explored a wide range of ML models for $f$, including MLP, SVM, and XGBoost~\cite{chen2016xgboost}, etc. We ultimately adopt XGBoost, a gradient tree model for the power model of all primitives, which yields the best accuracy and training time during validation. The benefits of our method include: 1) efficient to generate training data, i.e., 30 minutes for training set creation and minutes to train, and 2) primitive models can combine and generalize well to full designs.

\section{Experimental Results}

\subsection{Experimental Setup}

\begin{figure}[b!]
\vspace{-.35in}
  \centering
  \includegraphics[width=1\linewidth]{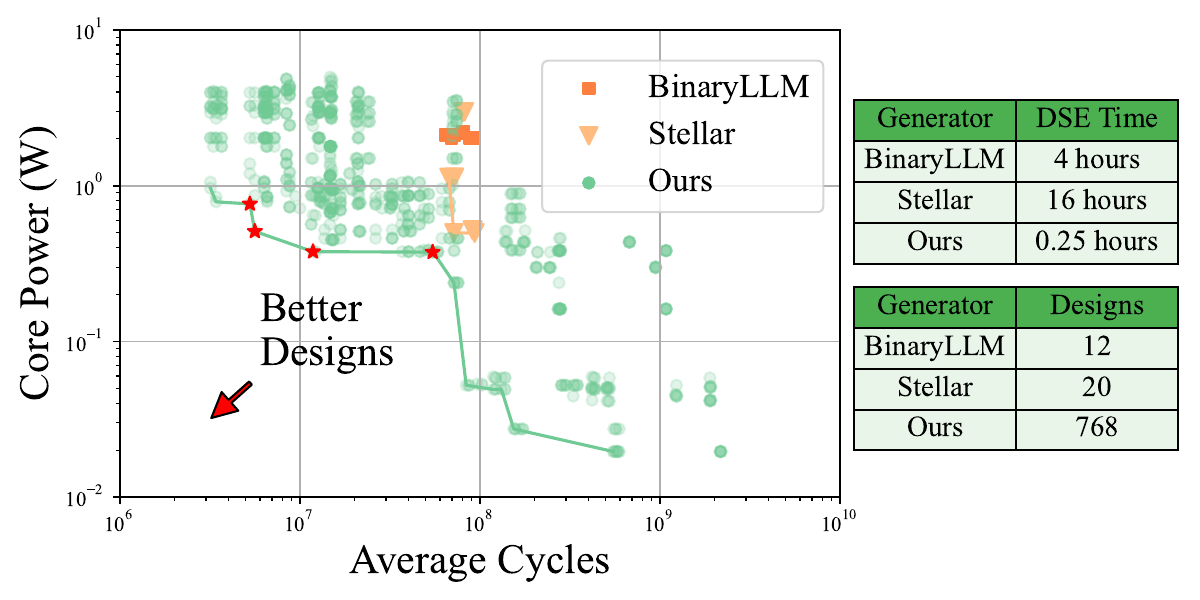}
   \vspace{-.3in}
  \caption{ Design space exploration comparison of hardware generators. Starred designs are our Pareto-optimal designs, which are further studied in Fig.~\ref{fig:dense} and Table~\ref{tbl:ppa}. }
  \label{fig:DSE}
 \vspace{-.1in}
\end{figure}
We evaluate our generator on several common LLM kernels and full networks. For benchmarking, we evaluate Bert-Large~\cite{Bert} and LLama~\cite{Llama}, which uses MHA, and Mistral~\cite{Mistral} and Qwen3-small~\cite{Qwen}, which uses GQA. Each model is evaluated in both the prefill and decode stages for different token lengths. 

\begin{figure}[t!]
 \vspace{-.55in}
  \centering
  \includegraphics[width=1\linewidth]{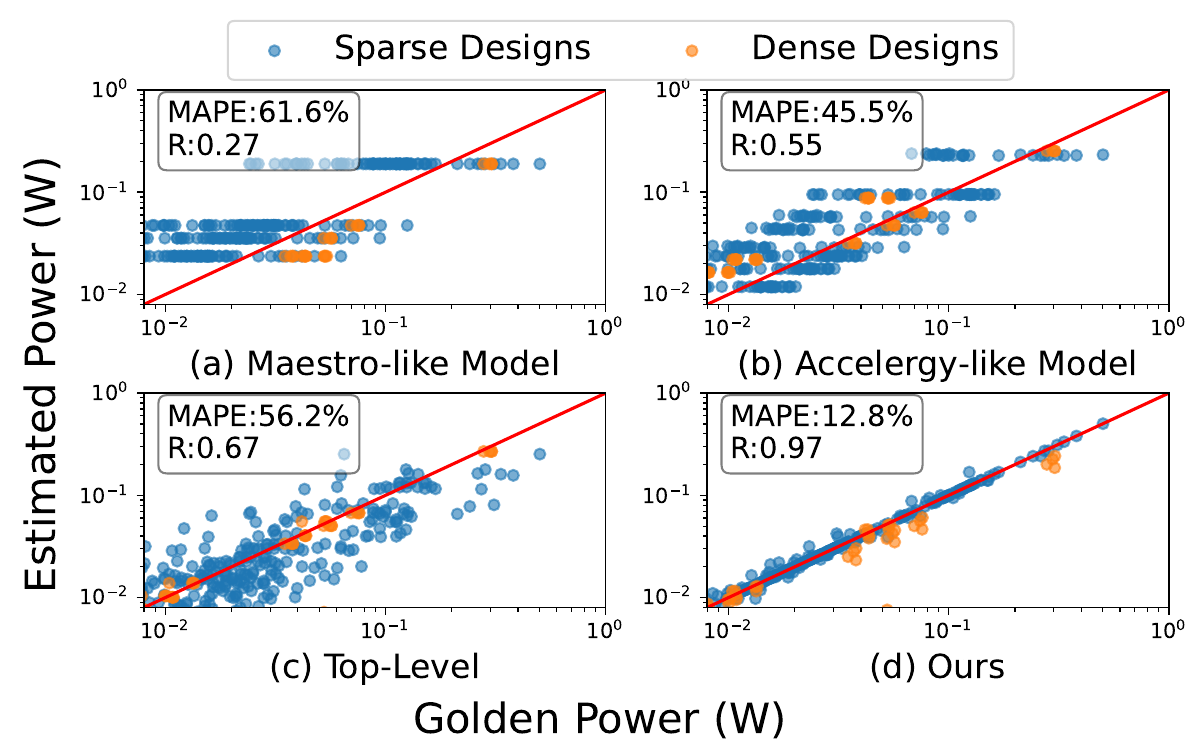}
   \vspace{-.3in}
  \caption{Comparison of early-stage power modelling against baselines Maestro~\cite{2019Understanding}, Accelergy~\cite{2019Accelergy}, and a top-level ML regressor. Our models are at the primitive-level. }
  \label{fig:PowerModel}
 \vspace{-.2in}
\end{figure}

We use Chisel~\cite{Chisel} to generate Verilog RTL for LLM accelerators. For ASIC generation and implementation, we use Synopsys Design Compiler for synthesis and area estimation under the TSMC 40nm technology node, Primetime and PTPX for power and timing analysis, and VCS for gate simulation. For power modelling analysis, the results are shown in Fig.~\ref{fig:PowerModel}, and we use the same setup and use Scipy for machine learning algorithms.

As shown in Fig.~\ref{fig:dense},~\ref{fig:DSE} and Table~\ref{tbl:ppa}, we conduct a DSE to benchmark and compare the performance and PPA of our \TITLE designs against prior hardware generators. We perform a grid-search over the combinations of 3 loop orders, 8 tilings, and 2 compute-value, 2 compute-group, and 4 memory-based sparsity mappings as described in Sec.~\ref{sec:sparity}, yielding 768 total designs. The baselines BinaryLLM~\cite{BinaryLLM} and Stellar~\cite{Stellar} are configured based on dataflow, which yield 12 and 20 distinct dataflows, respectively. The DSE's benchmarks are over single MHA layers. Each hardware is constrained to INT8 inference, connected to an 8 GB/s off-chip memory, assigned a maximum of 16KB per tensor and has a maximum of 8 TOPs and a clock speed of 1GHz.

\subsection{Generated Design Performance Results}
\label{sec:generateddesign}


Fig.~\ref{fig:dense} compares the performance of optimized designs. \TITLE's designs are named by their dataflow and sparse mappings. For example, design $D1$ is output-stationary (OS) with balanced tiling over $b,n,h,m$ and supports data sparsity $\Phi_x,w$, memory sparsity $\Phi_{mx,mw}$, blocked attention sparsity $\Phi_{g}$, and window attention $\Phi_{win}$.

Our designs are faster than designs from prior generators~\cite{Stellar,BinaryLLM}, because we support larger design spaces and fused operator dataflows, which greatly reduce memory accesses. 
For the prefill stage, our designs achieve up to 120k token/s (tps). Generally, larger LLMs, longer sequence lengths reduce performance due to quadratically increasing computation. GQA-based networks have even higher performance due to the reduced KV cache. 

For the decode stage on MHA benchmarks, our designs achieve up to 110 tps. GQA greatly reduce the kV cache by sharing heads, thereby greatly boosting the decode performance. Some designs are worse in decode because of lowered utilization especially over the input token length iterator $\vec{n}$.

Table~\ref{tbl:ppa} shows each design's detailed PPA metrics. Our designs have better overall PPA as highlighted by the figure of merit (FoM), which is 58$\times$ higher than prior art. Our design's performance, area, power, and utilization are more optimal. The maximum clock is slightly higher for prior works because of hardware simplicity. Design D1 achieves similar performance as prior works but consumes less power and has better utilization due to the better sparsity support. Design D2 has more balanced tilings, at the cost of slightly larger networks, resulting in lower clock frequency but improved performance. Designs D3 and D4 use larger PE arrays, resulting in improved performance but at the cost of energy and area. We observe that the sparse mappings $\Phi_g,\Phi_{win}$ greatly contribute to improving our model's FoM compared with prior works, which have limited support for different types of sparsity.

\subsection{Design Space Exploration Results}
\label{sec:dseresults}

As shown in Fig.~\ref{fig:DSE}, our generator yields a more optimal design space and better Pareto-optimal curve, due to our larger space of sparsity and dataflow support. At a 700mW constraint, our designs have 10$\times$ better performance, and at the same cycle count of 8$\times10^7$ cycles, our designs have 1.4$\times$ better power.  Because BinaryLLM designs are not sparse, the designs have higher latency. Stellar designs support only compute and memory value sparsity and is not comprehensive. The DSE runtime and space are also better than prior methods due to our early-stage model.

\subsection{Early-Stage Power Modelling Results}
\label{sec:powermodel}

Fig.~\ref{fig:PowerModel} shows our early-stage power models and comparison against Maestro~\cite{2019Understanding}, Accelergy~\cite{2019Accelergy}, and a top-level version of our power model. Our model is based on primitives, which are trained by varying the primitive type and port data and collecting golden post-synthesis power labels. Each primitive has around 256 training data points, requires around 30 minutes for data collection, and is trained in seconds. For validation, we use \TITLE to generate around 20 dense and 200 sparse designs by varying the dataflows, sparsity set and tilings. Our power models achieve 12.8$\%$ error and $R=0.97$, which are better calibrated compared to prior art. Prior art can not capture power accurately because they lack 1) detailed toggling features, and 2) accurate models for high-energy and data-sensitive components, such as sparse networks.

\subsection{Sparsity Set Design Space Tradeoffs}


\begin{figure}[t!]
  \centering
   \vspace{-.55in}
  \includegraphics[width=1\linewidth]{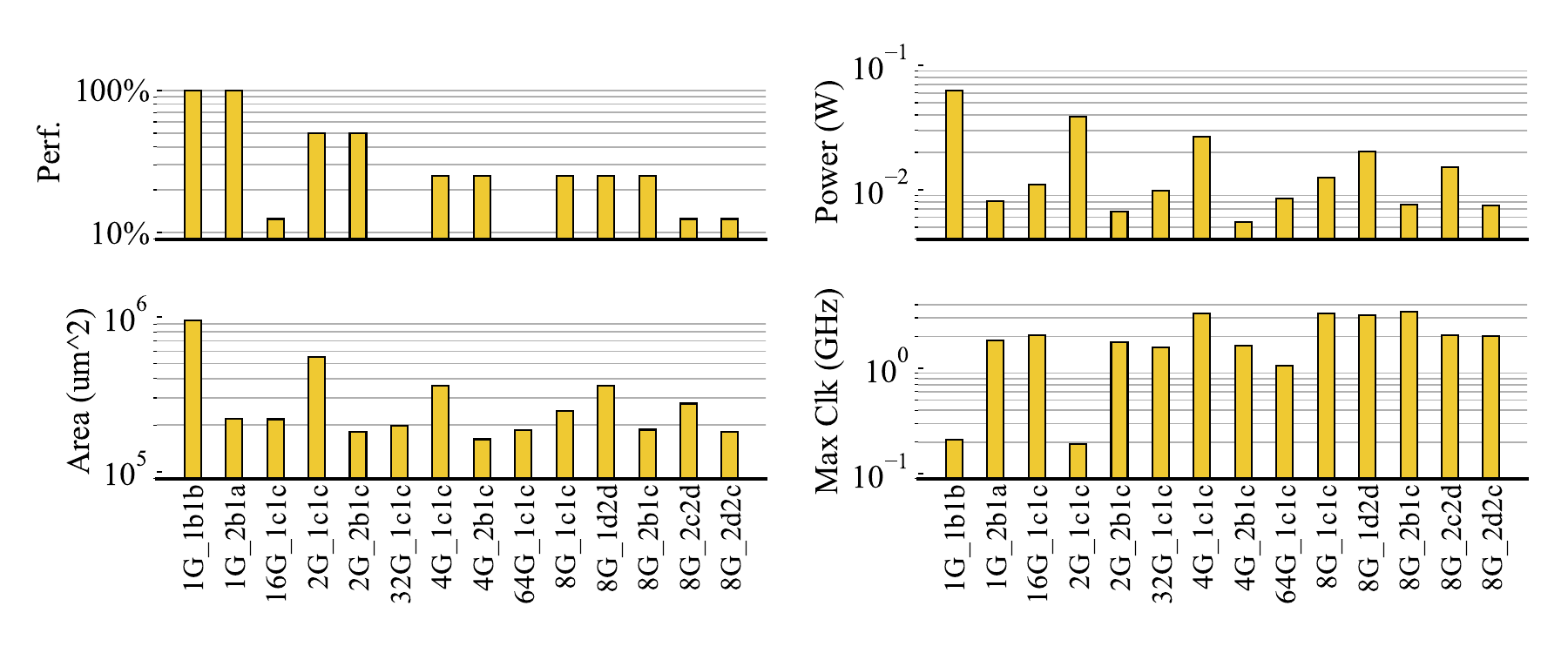}
   \vspace{-.4in}
  \caption{ Comparison of sparse networks for attention. The design space corresponds to the structures in Fig.~\ref{fig:valsparse}.  }
  \label{fig:SparsityTradeoffs}
 \vspace{-.2in}
\end{figure}

To better understand the sparse networks supported by \TITLE, we generate a variety of 8$\times$8 arrays for attention with $Q=\{(x==0), (w==0)\}$. Referring to Fig.~\ref{fig:SparsityTradeoffs}, a design with name $2G\_2b1a$ means it has group $g=2$ and $x,w$ networks that are $2b$ and $1a$, respectively. The PPA varies greatly. Unicast networks (1b) have the highest performance but higher delays and power due to large crossbars. Multicast networks (2b) save power and area by broadcasting, but have reduced performance due to under-utilization, especially for low-batch inferences. Grouped networks (1c) improve max clock but trade off performance due to network imbalance. A mixture of networks, such as $2b1a$, yields a good overall PPA.

\section{  Conclusion }

  In this paper, we propose \textit{\TITLE}, an AI chip generator and early-stage estimator for LLM applications. Our framework explores the vast space of sparse and fused-operator dataflows for LLM applications. Using our early-stage PPA ML-based models, we find optimal designs better than prior generators. We identified several designs and synthesized them into ASIC designs. We find that our reconfigurable design, coupled with sparse networks, balanced tilings and sparse sets focusing on compute-group and memory-based sparsity, greatly helps in all stages of LLM acceleration.
  


\begin{acks}
This work is supported by Hong Kong Research Grants Council (RGC) CRF-YCRG C6003-24Y, National Natural Science Foundation of China (NSFC) 62304192. It was partially conducted by ACCESS – AI Chip Center for Emerging Smart Systems, supported by the InnoHK initiative of the Innovation and Technology Commission of the Hong Kong Special Administrative Region Government. 
    
\end{acks}

\bibliographystyle{ACM-Reference-Format}
\bibliography{ref}

\end{document}